\documentclass[onecolumn,floatfix,aps,superscriptaddress]{revtex4}
\usepackage{graphicx}
\usepackage{amsmath}
\usepackage[colorlinks=true, citecolor=blue, urlcolor=blue, linkcolor = blue]{hyperref}
\usepackage{amssymb}
\usepackage{bm}
\usepackage{color}
\usepackage{bookmark}
\usepackage{tabularx}
\usepackage{mathtools}
\usepackage{microtype}
\usepackage{cancel}
\usepackage{relsize}

\usepackage{xcolor}
\begin{document}

\title{Bending of  solitonic beams modelled by coupled KMN equation in optical birefringent fibers}

\author{Abhik Mukherjee}
 \affiliation{Fundamental Quantum Science Division,
 Tsung-Dao Lee Institute,
 Shanghai Jiao Tong University, Shanghai,  Republic of China}

  % (includes institutions and visitors)
%\date{\today}

\begin{abstract}
The dynamics of $(2+1)$ dimensional optical solitonic beams in birefringent fibers without four wave mixing terms are considered using the recently introduced 
 coupled Kundu Mukherjee Naskar (KMN) equation.  The arbitrary bending of solitonic beams in this coupled optical fiber system has been discussed with various two and three dimensional plots. Such analytical results on the bending of optical pulse may pave new directions of research in this  field.
\end{abstract}

\pacs{}
\maketitle

\section{Introduction}
 The upsurge of research   in nonlinear fiber optics
 has started in early seventies leading extensive research on Raman and Brillouin Scattering, optically induced birefringence, parametric four-wave mixing, self phase modulation etc. \cite{Agarwal}. An important contribution was made when it was suggested that optical fibers can support soliton like pulses which was experimentally verified in 1980 \cite{Agarwal}.  
 After its discovery in nonlinear optics, the dynamics of soliton pulses has been   investigated extensively in (1+1) and higher dimensions using both numerical and analytical techniques. The main specialty of soliton, which is  the exact solution of integrable systems, is its stability \cite{Agarwal}. The amplitude and velocity of solitons remain constant during its propagation. On the other hand, the presence of lossy or time dependent terms in the evolution equation makes it non-integrable causing the velocity of solitary waves to change.   Recently,  a deep research interest has been developed on some special intricate exact solutions of nonlinear systems  like accelerating solitons \cite{accsol, Kundu-PLA-2019, Zaqilao},
 topological solitons \cite{toposol, TopB1, TopB2, TopB3, TopB4}, rogue wave solutions  \cite{ourpaper, Rogue} etc.

In the last decade, a  new completely integrable (2+1) dimensional nonlinear evolution equation was derived from the basic equations of hydrodynamics by Anjan Kundu, Abhik Mukherjee and Tapan Naskar to describe the dynamics of two-dimensional oceanic rogue wave phenomena \cite{ourpaper}. This  equation was also re-derived in magnetized plasma system \cite{Kundu-PoP-2015} to model nonlinear ion acoustic waves.  The detailed exploration of integrable properties and soliton solutions of the equation were also carried out \cite{ourpaper, Kundu-PoP-2015, Radu}.
The equation was derived in nonlinear optical system  by Kundu and Naskar in \cite{KunduNaskar} to explain the arbitrary bending phenomena of optical solitonic beam. Such bending is controlled by the boundary population inversion of atoms in an Erbium doped medium.
A first order rogue wave solution of this equation was obtained by one fold Darboux Transformation  in \cite{CNSNS-He-2016} where  the authors have named this new equation as Kundu-Mukherjee-Naskar (KMN) equation. Thereafter, an extensive amount of research has been
done on KMN equation exploring its various
interesting properties. Many kinds of solutions
of KMN equation were derived like higher order rational solutions \cite{Roman-2017}, optical dromions and domain walls \cite{Biswas1-2020, Gaxiola}, various  optical soliton solutions  having different features \cite{CJP-2019,Optik-2019a,Optik-2019b,Rizvi,Sulaiman,Sudhir1,Sudhir2, Aliyu, Hadi, Bilal, Triki, Zayed}, different analytic  solutions \cite{Jalili,Khater,Kumar, Dang, Salam,Baskonus, Ren, Kudryashov}, power series solutions  \cite{MPLB-2018}, complex wave solutions \cite{Jhangeer, Rodica}, periodic solutions (via variational principle) \cite{J-He}, solitons in birefringent fiber system \cite{Optik-2KMN-2019,Optik-2KMN-2019a,Yildirim-f} etc. 
Integrable properties such as  Lax pair, conservation laws, higher order soliton solutions via Hirota method, symmetry analysis, nonlinear self-adjointness property etc. of KMN equation has been explored in \cite{ourpaper,Sudhir2,Radu,Rodica,Sachin}. 
Time fractional KMN equation has been introduced in \cite{He-Dib} to explore periodic properties.

One of the important properties of KMN equation is that it supports both bright as well as dark soliton solutions \cite{Sudhir1} that actually comes from the current like nonlinearity (which depends on both wave envelope and it's spatial derivative) present in the equation. This property  is different from the standard Nonlinear Schrodinger (NLS) equation which admits only bright and dark soliton solution for focusing and defocusing nonlinearities respectively. 
In real  experimental conditions, the localized wave  structures  may have various shapes depending on the physical system. Hence, the search for exact solitary wave solutions with variable amplitude or velocity is also an important problem to investigate.  In \cite{MyOptik},  we have derived a special lump, snoidal wave and topological soliton solution that can get curved in $x-y$ plane arbitrarily due to presence of an arbitrary function of space($x$) and time($t$) in their analytic expressions.  Thus, this important bending feature  which comes  from the Galilean co-variance property and current like nonlinearity present in KMN equation, is rare for a completely integrable equation having constant coefficients. 
 
 The interaction of two  modes of shorter wavelengths in optical fiber system has been discussed  using a system of coupled Nonlinear Schrodinger Equations  in (1+1) dimensions
  \cite{Radhakrishnan, Radhakrishnan2, Radhakrishnan3} as well as in higher dimensions \cite{Liu2}. It is also important to state that 
 the single modes in the optical fiber  are actually bimodal due to
the presence of birefringence \cite{Menyuk}.
 The coupled KMN equation was introduced in \cite{Optik-2KMN-2019, Optik-2KMN-2019a} to model (2+1) dimensional optical solitons in birefringent fiber without four wave mixing terms. In this paper, our aim is to explore the bending phenomena of optical solitonic beams in coupled KMN system using Hirota bilinearization method which has not been explored till now.
 Arbitrary bending light has been discussed  in \cite{KunduNaskar} where the boundary value of population inversion of the medium becomes non uniform. On the other hand, we will show that the bending property for coupled KMN system is intrinsic i.e, does not depend on any external function.  

The paper is outlined as follows: the coupled KMN equation is introduced in section-II and its Galilean co-variance is derived in section-III. Exact bright and dark solitons are derived using Hirota method in section IV and V respectively. The bending of optical solitonic beams is discussed in section-VI. Conclusive remarks are stated in section-VII followed by acknowledgments and bibliography.

\section{Coupled KMN equation}
The Kundu Mukherjee Naskar (KMN) equation which was first derived in modelling oceanic rogue waves \cite{ourpaper} is given as 
\begin{equation}
 i u_t + u_{xy} + 2 i u (u u_x^{*} - u^{*}u_x) = 0, \label{KMN}
\end{equation}
where the wave profile is given by  $u(x,y,t)$  and the subscripts denote partial derivatives w. r. to space (x, y) and time (t) coordinates. The asterisk denotes complex conjugation. The coupled Kundu Mukherjee Naskar (KMN) equation was first introduced by Yildirim \cite{Optik-2KMN-2019, Optik-2KMN-2019a} in modelling the dynamics of (2+1) dimensional optical solitonic beams in birefringent fibers without four wave mixing terms. In this work, we take the same system of equations taken in \cite{Optik-2KMN-2019, Optik-2KMN-2019a} with  $a_i = 1, b_i = c_i = 2$, $i = 1, 2$  for simplicity which can be written as :   
\begin{eqnarray}
 &{}iu_t + u_{xy} + 2 i u \{  (u u_{x}^* - u^*u_{x}) +  (v v_{x}^* - v^*v_{x})\} = 0, \nonumber \\
 &{}iv_t +  v_{xy} + 2 i v \{  (v v_{x}^* - v^*v_{x}) +  (u u_{x}^* - u^*u_{x})\} = 0, \label{cKMN}
\end{eqnarray}
where $u(x,y,t), v(x,y,t)$ are two wave profiles. Due to the presence of the current like nonlinearities in Eq. (\ref{cKMN}), it shows few interesting features which will be discussed below.

 \section{Galilean co-variance (GC)}
 The Galilean Co-variance (GC) property of KMN equation has been derived in \cite{MyOptik}. Generalizing the same mathematical procedure to (\ref{cKMN}), we can also find the GC of coupled KMN equation.
 The equations (\ref{cKMN}) remains co-variant as 
\begin{eqnarray}
 &{}iU_T +  U_{XY} + 2 i U \{  (U U_{X}^* - U^*U_{X}) +  (V V_{X}^* - V^*V_{X})\} = 0, \nonumber \\
 &{}iV_T +  V_{XY} + 2 i V \{  (V V_{X}^* - V^*V_{X}) +  (U U_{X}^* - U^*U_{X})\} = 0, \label{cKMN2}
\end{eqnarray}
 under the Galilean transformation:
 \begin{eqnarray}
  X = x - a t, \ \
  Y = y, \ \
  T = t, \ \
  u(x, y, t) = U(X, Y, T) \ e^{i a Y}, 
  v(x, y, t) = V(X, Y, T) \ e^{i a Y},\label{GT}
 \end{eqnarray}
where $a$ is the constant frame velocity. This co-variance property is one of the crucial feature for the bending of optical solitonic beams which will be discussed below.

\section{ bright Solitons }
Optical solitons of the coupled KMN equation (\ref{cKMN}) has been derived in \cite{Optik-2KMN-2019, Optik-2KMN-2019a} using trial equation approach and
modified simple equation approach. Now using Hirota bilinearization method, we will derive one bright and one dark solitons of (\ref{cKMN}) which has not been done before as far as our knowledge goes.
The Hirota bilinear transformation \cite{Radhakrishnan,Radhakrishnan2,Radhakrishnan3} is defined as
\begin{equation}
 u(x,y,t) = \frac{g(x,y,t)}{f(x,y,t)}, \ \ \  v(x,y,t) = \frac{h(x,y,t)}{f(x,y,t)}, \label{HT}
\end{equation}
where $g, h$ are complex functions and $f$ is a real function. Using (\ref{HT}) in (\ref{cKMN}) we can form the following bilinear equations :
\begin{eqnarray}
&{} (i D_t +  D_x D_y)(g . f) = 0, \nonumber \\
&{} (i D_t +  D_x D_y)(h . f) = 0, \nonumber \\
&{}  D_x (f . f_y) + i  D_x (g^{*} . g) + i  D_x (h^{*} . h) = 0, 
% &{}  D_x (f . f_y) + i  D_x (h^{*} . h) + i  D_x (g^{*} . g) = 0, 
\label{HBform}
\end{eqnarray}
where the Hirota bilinear  operator \cite{Radhakrishnan} is defined as
\begin{eqnarray}
 &{} D_x^m D_t^n \  g(x,t) \ . \  f(x,t) = (\frac{\partial}{\partial x} - \frac{\partial}{\partial x'})^m
 \ (\frac{\partial}{\partial t} - \frac{\partial}{\partial t'})^n  \ g(x,t) \ f(x',t')|_{x=x',t=t'}. \label{DOp}
\end{eqnarray}
 Now we will proceed in the standard way to find the bright and dark soliton solutions. In order to find one bright soliton, we assume
\begin{eqnarray}
 &{} g = \epsilon g_1, \ \ h = \epsilon h_1, \ \ f = 1 + \epsilon^2 f_2, 
 \label{expan}
\end{eqnarray}
where $\epsilon$ is the standard expansion parameter and $g_1, h_1, f_2$ are new variables.
Now using (\ref{expan})
in (\ref{HBform}), we can get the following differential equations at different orders of $\epsilon$ as 
\begin{eqnarray}
&{}O(\epsilon) \ : \ (i D_t +  D_x D_y)(g_1 . 1) = 0, \
(i D_t +  D_x D_y)(h_1 . 1) = 0, \nonumber \\
&{}O(\epsilon^2) \ : \ - f_{2xy} + i  D_x (g_1^{*} . g_1) + i  D_x (h_1^{*} . h_1) = 0, \
 \nonumber \\
&{}O(\epsilon^3) \ : \ (i D_t +  D_x D_y)(g_1 . f_2) = 0, \
(i D_t +  D_x D_y)(h_1 . f_2) = 0, \nonumber \\
&{}O(\epsilon^4) \ : \
 D_x (f_2 . f_{2y}) = 0.
\label{HBform1}
\end{eqnarray}
Solving the systems of equations (\ref{HBform1}) one can find the solutions as 
\begin{eqnarray}
 &{}g_1(x,y,t) = \alpha \  e^{\eta_1}, h_1(x,y,t) =  \beta \ e^{\eta_1}, \ \eta_1(x,y,t) = k \ x + m \ y + i \ k \ m \ t + \eta_0, \nonumber \\
 &{}f_2(x,y,t) = \{\frac{i \  (k^{*} - k)}{(k^{*} + k)(m^{*} + m)}\}\{|\alpha|^2 + |\beta|^2 \} \ e^{(\eta_1 + \eta_1^{*})}, \ e^{\delta_1} = \epsilon^2
 \frac{k_I (|\alpha|^2 + |\beta|^2)}{2 k_R m_R}
 \end{eqnarray}
where $k, m, \eta_0, \alpha, \beta$ are complex constants. Simplifying above expressions, we can obtain the analytical expressions of two modes as 
\begin{eqnarray}
&{}u =  \frac{\epsilon g_1}{1+\epsilon^2 f_2}
= \mu_1 \ Sech[k_R x + m_R y - (k_R m_I + k_I m_R) t ] \ e^{i [k_I x + m_I y + (k_R m_R - k_I m_I)t] }, \nonumber \\
&{}v =  \frac{\epsilon h_1}{1+\epsilon^2 f_2}
= \mu_2 \ Sech[k_R x + m_R y - (k_R m_I + k_I m_R) t ] \ e^{i [k_I x + m_I y + (k_R m_R - k_I m_I)t ]}, \nonumber \\
{where}, 
&{} \mu_1 = \frac{\alpha}{2} \sqrt{\frac{2 k_R m_R}{k_I(|\alpha|^2 + |\beta|^2)}},  \ \mu_2 = \frac{\beta}{2} \sqrt{\frac{2 k_R m_R}{k_I(|\alpha|^2 + |\beta|^2)}}, \alpha = |\alpha|e^{i \theta}, \ \beta = |\beta|e^{i \phi}
\label{bsol}
\end{eqnarray}
and ``$R$'' and ``$I$'' in the subscripts  denote the real and imaginary parts of the complex constants. In the above simplification, we have chosen the arbitrary constants $\eta_{0R}, \eta_{0I}, \phi, \theta, \delta_1$  in such a way that they get cancelled and we get the expressions (\ref{bsol}). The 3 - dimensional plots of the two bright solitonic modes $u$ and $v$ are shown in FIG.1.

\begin{figure}[hbt!]
\centering
\includegraphics[width=7cm]{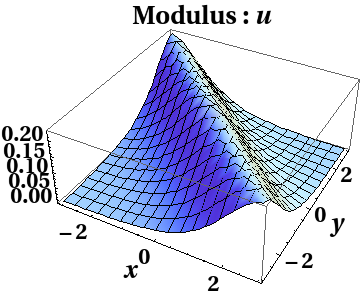}
\ \
\includegraphics[width=7
cm]{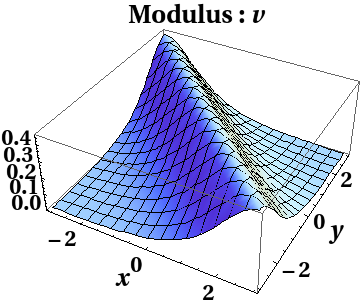}
\caption{The modulus of the two optical bright solitonic modes $u,v$ respectively are plotted in $x-y$ plane at $t=0$. The constants are chosen as $k_R = 1,
k_I = 2,
m_R = 1,
m_I = 2, |\alpha| = 1, |\beta| = 2.$} 

\label{NDTEplot}
\end{figure}

\section{Dark Solitons}
The importance of the KMN model lies in the fact that it supports both  bright and dark soliton solutions. In case of NLSE with Kerr like nonlinearity, bright and dark solitons occur for focusing and defocusing nonlinearities respectively. In order to find one dark solitons of (\ref{cKMN}) we assume
\begin{eqnarray}
&{}g = g_0(1 + \epsilon g_1), \ h = h_0(1 +  \epsilon h_1), \ f = 1 + \epsilon f_1, \label{expan2}   
\end{eqnarray}
where $\epsilon$ is the same expansion parameter as discussed before. Thus using this expansion (\ref{expan2}) in (\ref{HBform}) we would get the following differential equations at different orders of $\epsilon$.

At $O(\epsilon^0)$, we would get
\begin{eqnarray}
&{}(i D_t +  D_x D_y)(g_0 . 1) = 0, \
(i D_t +  D_x D_y)(h_0 . 1) = 0, \  D_x (g_0^{*} . g_0) +   D_x (h_0^{*} . h_0) = 0. 
 \label{0order}
\end{eqnarray}
In order to solve (\ref{0order}), we assume 
\begin{equation}
 g_0 = \tau_1 e^{i \psi_1}, h_0 = \tau_2 e^{i \psi_2}, \psi_i = k_i x+ m_i y+ n_i t + \psi_i^{0}, 
\end{equation}
with $i=1,2$, where $\psi_i$s are real and $\tau_i$s are complex quantities. Solving (\ref{0order}), we would get 
\begin{eqnarray}
&{}n_1 = - k_1 m_1, \ n_2 = -k_2 m_2, \ k_1 |\tau_1|^2 + k_2 |\tau_2|^2 = 0. \label{relconst}
\end{eqnarray}
At $O(\epsilon)$, we would get the following differential equations :
\begin{eqnarray}
&{}(g_{1t}-f_{1t}) + k_1 (g_{1y} - f_{1y}) + m_1 (g_{1x} - f_{1x}) - i(g_{1xy} + f_{1xy}) = 0, \nonumber \\
&{}(h_{1t}-f_{1t}) + k_2 (h_{1y} - f_{1y}) + m_2 (h_{1x} - f_{1x}) - i(h_{1xy} + f_{1xy}) = 0, \nonumber \\
&{}f_{1xy} = 2 k_1 |\tau_1|^2 (g_1 + g_1^{*}-h_1 - h_1^{*}) - i |\tau_1|^2(g_{1x}-g_{1x}^{*})-i|\tau_2|^2(h_{1x} - h_{1x}^{*}).
 \label{1order}
\end{eqnarray}
Equations (\ref{1order}) admit the following solutions :
\begin{eqnarray}
&{} g_1 = Z_g e^{\xi_1}, \ h_1 = Z_h e^{\xi_1}, \ f_1 = e^{\xi_1}, \
&{} \xi_1 = P_1 x + Q_1 y + R_1 t + \xi_1^{0},
\end{eqnarray}
where $Z_g, Z_h$ are complex constants and $P_1, Q_1, R_1, \xi_1^0$ are real constants that are related by following constraints :
\begin{eqnarray}
 &{}Z_g = \frac{[(R_1 + k_1 Q_1 + m_1 P_1) + i P_1 Q_1]^2}{[(R_1 + k_1 Q_1 + m_1 P_1)^2 + P_1^2 Q_1^2]}, \
 Z_h = \frac{[(R_1 + k_2 Q_1 + m_2 P_1) + i P_1 Q_1]^2}{[(R_1 + k_2 Q_1 + m_2 P_1)^2 + P_1^2 Q_1^2]}, \nonumber \\
 &{} Q_1 = (\frac{2k_1}{P_1})|\tau_1|^2(Z_g + Z_g^{*}-Z_h - Z_h^{*}) - i|\tau_1|^2(Z_g - Z_g^{*})- i|\tau_2|^2(Z_h - Z_h^{*}). \label{RelConstants}
\end{eqnarray}
Absorbing $\epsilon$ in $e^{\xi_1}$, the one dark solitons can be expressed as
\begin{eqnarray}
 &{} u = (\tau_1/2) e^{i \psi_1}[(1 + Z_g)-(1 - Z_g) \tanh{(\xi_1/2)}], \nonumber \\
 &{}v = (\tau_2/2) e^{i \psi_2}[(1 + Z_h)-(1 - Z_h) \tanh{(\xi_1/2)}].\label{uv}
\end{eqnarray}

The 3 - dimensional plots of the two dark solitonic modes $u$ and $v$ are shown in FIG.2.

\begin{figure}[hbt!]
\centering
\includegraphics[width=7cm]{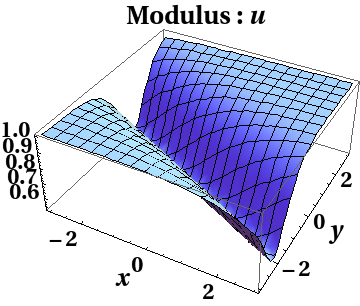}
\ \
\includegraphics[width=7
cm]{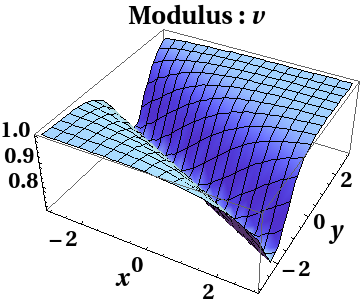}
\caption{The modulus of the two optical dark solitonic modes $u,v$ respectively are plotted in $x-y$ plane at $t=0$. The constants are chosen as $k_1 = 1,
k_2 = -1, P_1 = 1, Q_1 = 2, R_1 =2, m_2 = 2, m_1 = \sqrt{5}-5, 
\tau_1 = \frac{1 + i}{\sqrt{2}}, \tau_2 = \frac{1 - i}{\sqrt{2}}$
.} 

\label{NDTEplot}
\end{figure}

\section{Bending of optical solitonic beams}

One important property of KMN model (\ref{KMN}) is that it processes curved solitons that is explored in \cite{MyOptik}. Due to the presence of the current like nonlinearity and Galilean co-variance in (\ref{KMN}), this special dark soliton can get curved arbitrarily in $x-y$ plane. In \cite{MyOptik}, one curved soliton has been derived using ansatz approach. Now, we will rederive such curved one soliton solution of Eq. (\ref{KMN}) using Hirota bilinearization method that has not been done before.

\subsection{ Derivation of one curved dark soliton solution of $KMN$ equation (\ref{KMN}) via Hirota bilinearization method}
Here, we will derive  one curved dark soliton solution of $KMN$ equation (\ref{KMN}) via Hirota bilinearization technique. We consider the KMN equation in the Galilean frame $(X - Y - T)$ that is discussed in section - III as 
\begin{eqnarray}
 &{} i U_T + U_{XY} + 2 i U (U U_X^{*} - U^{*} U_X) = 0, \label{KMNG}
\end{eqnarray}
where $X = x-at, \ Y = y, T = t, \  u(x,y,t) = U(X,Y,T) e^{i a Y}.$
We use the bilinear transformation as 
\begin{equation}
 U(X, Y, T) = \frac{G(X, Y, T)}{F(X, Y, T)}, 
\end{equation}
where $G, F$ are complex and real functions respectively, to get the pair of bilinear equations :
\begin{eqnarray}
&{} [i D_T +  D_X D_Y] \ (G   . F) = 0, \nonumber \\
&{}  D_X \ [ (F . F_Y) + i  \ (G^{*} . G) ] = 0.
 \label{HBformKMN3}
\end{eqnarray}

In order to find one dark curved soliton of (\ref{KMNG}) we assume
\begin{eqnarray}
&{}G = G_0(1 + \epsilon G_1),  \ F = 1 + \epsilon F_1, \label{expan3}   
\end{eqnarray}
where $\epsilon$ is the same expansion parameter as discussed before.
At this stage, we would assume that $G, F$ are independent of time $T$. That will not make the solution static, the reason will be explored later.
Thus using this expansion (\ref{expan3}) in (\ref{HBformKMN3}) we would get the following differential equations at different orders of $\epsilon$.

At $O(\epsilon^0)$, we would get
\begin{eqnarray}
&{} G_{0XY} = 0, \ (G_{0X}^{*}G_0 - G_0^{*}G_{0X}) = 0
 \label{0KMNG}
\end{eqnarray}
In order to solve (\ref{0KMNG}), we assume 
\begin{equation}
 G_0 = \tau e^{i \psi},  \psi = k X+ m Y + \psi^{0}, 
\end{equation}
where $k, m, \psi^0$ are real constants and $\tau$ is complex constant. Solving (\ref{0KMNG}), we would get $k=0$. The the solution is  
\begin{eqnarray}
&{} G_0 = \tau e^{i (m Y + \psi_0)}.
\end{eqnarray}
At $O(\epsilon)$, we would get 
\begin{eqnarray}
&{}G_{1XY} + F_{1XY} + im (G_{1X} - F_{1X}) = 0, \nonumber \\
&{} F_{1XY} = i |G_0|^2 (G_{1X}^{*} - G_{1X}).
 \label{1KMNG}
\end{eqnarray}
Equations (\ref{1KMNG}) admit the following solutions :
\begin{eqnarray}
&{} G_1 = Z_g e^{\xi}, \  \ F_1 = e^{\xi}, \
&{} \xi = A(X) + m_1 Y +  \xi^{0},
\end{eqnarray}
where $Z_g$ is complex constant and $m_1,  \xi^{0}$ are real constants. We have introduced an arbitrary real function $A(X)$ which will ultimately cause the bending of solitons. The constants are related as 
\begin{eqnarray}
 &{}Z_g = \frac{i m - m_1}{i m + m_1}, \ |\tau|^2 = \frac{(m_1^2 + m^2)}{4m}.\label{RelCKMNG}
\end{eqnarray}
 If the function $A(X)$ is nonlinear in $X$, the soliton will bend in $X-Y$ plane making curvature. The final soliton solution can be written as 
\begin{eqnarray}
 &{} U = \frac{\tau}{2}[(1 + Z_g) - (1-Z_g) \tanh{(\xi /2)}] e^{i (m Y)}. 
\end{eqnarray}
Now moving back to the original frame $x-y-t$ and old variable $u$, we would get
\begin{eqnarray}
 &{} u = \frac{\tau}{2}[(1 + Z_g) - (1-Z_g) \tanh{(\xi /2)}] e^{i (m y + a y)}, \xi = A(X) + m_1 y, \  X = x - a t. \nonumber \\
 &{}Z_g = \frac{i m - m_1}{i m + m_1}, \  \tau = (m + i m_1)/\sqrt{4m}. 
\end{eqnarray}
Hence the soliton bends due to the presence of $A(X),$ which is a nonlinear function of $X = (x - a t).$ For the different nonlinear choices of $A(X), X = x- a t$, we would get different types of bending of solitons which is discussed in \cite{MyOptik}.

\subsection{Curved solitons for coupled $KMN$ equations}

Now we will concentrate on the main findings of this work. Similar to the previous case, we  consider the coupled KMN system  (\ref{cKMN2}) in the Galilean frame $X-Y-T$
 and take $U = G/F, \ V = H/F$, where $H,G, F$ are independent of $T$. The functions $G, H$ are complex and $F$ is a real function.

Expanding as :
\begin{eqnarray}
&{}G = G_0(1 + \epsilon G_1), \ H = H_0(1 +  \epsilon H_1), \ f = 1 + \epsilon F_1, \label{expan2cKMN}   
\end{eqnarray}
where $\epsilon$ is the same expansion parameter as discussed before, we would get the following differential equations at different orders of $\epsilon$.

At $O(\epsilon^0)$, we would get
\begin{eqnarray}
&{} G_{0XY} = 0, \ H_{0XY} = 0, \  (G_{0X}^{*}G_0 - G_0^{*}G_{0X}) + (H_{0X}^{*}H_0 - H_0^{*}H_{0X})  = 0
 \label{0KMNcG}
 \end{eqnarray}
 with following solutions
\begin{eqnarray}
&{} G_0 = \tau_1 e^{i(m_1 Y + \psi_1^0)}, \ H_0 = \tau_2 e^{i(m_2 Y + \psi_2^0)},
\end{eqnarray}
where $m_{1,2}, \psi_{1,2}^0$ are real constants and $\tau_{1,2}$ are complex constants.

At $O(\epsilon)$, we would get the following differential equations :
\begin{eqnarray}
&{}G_{1XY} + F_{1XY} + i m_1 (G_{1X} - F_{1X}) = 0, \ H_{1XY} + F_{1XY} + i m_2(H_{1X} - F_{1X}) = 0 \nonumber \\
&{} F_{1XY} = i |\tau_1|^2 (G_{1X}^{*} - G_{1X}) + i |\tau_2|^2 (H_{1X}^{*} - H_{1X})
 \label{1cKMNG}
\end{eqnarray}
with the solutions :
\begin{eqnarray}
&{} G_1 = Z_g e^{\xi_1}, \ H_1 = Z_h e^{\xi_1}, \ F_1 = e^{\xi_1}, \
&{} \xi_1 = B(X) + P Y +  \xi_1^{0},
\end{eqnarray}
where $Z_g, Z_h$ are complex constants and $P, \xi_1^0$ are real constants that are related by following constraints :
\begin{eqnarray}
&{}Z_g = \frac{i m_1 - P}{i m_1 + P}, Z_h = \frac{i m_2 - P}{i m_2 + P}, \ P = 2(Z_{gI} |\tau_1|^2 + Z_{hI} |\tau_2|^2),
  \label{RelConcKMN}
\end{eqnarray}
where $I$ in the subscript denotes the imaginary part. The function $B(X)$ is the arbitrary function of $X = (x - a t).$ For different nonlinear choices of $B(X)$, we would get different kinds of bending of $u, v$. 
Absorbing $\epsilon$ in $\xi_1^0$, the one dark solitons in the old frame $x-y-t$ can be expressed as
 \begin{eqnarray}
  &{} u = \frac{\tau_{1}}{2}
   [(Z_g -1) \tanh{\frac{\xi_1}{2}} + (Z_g + 1)]
   e^{i (m_1 y + a y + \psi_{10})}, \
  v = \frac{\tau_{2}}{2}
   [(Z_h -1) \tanh{\frac{\xi_1}{2}} + (Z_h + 1)]
   e^{i (m_2 y + ay + \psi_{20})}, \nonumber \\
   &{} Z_g = \frac{i m_1 - P}{i m_1 + P}, \  Z_h = \frac{i m_2 - P}{i m_2 + P}, \ P = 2(Z_{gI} |\tau_1|^2 + Z_{hI} |\tau_2|^2), \  \xi_1 = B(X) + P Y +  \xi_1^{0}, X = x - a t.
   \label{csolfinal}
\end{eqnarray}
The bending of such dark solitons $u, v$ are shown as contour plots in $x-y$ plane for different choices of $B(X)$ in FIG. 3, 4, 5 respectively.

\begin{figure}[hbt!]
\centering
\includegraphics[width=7cm]{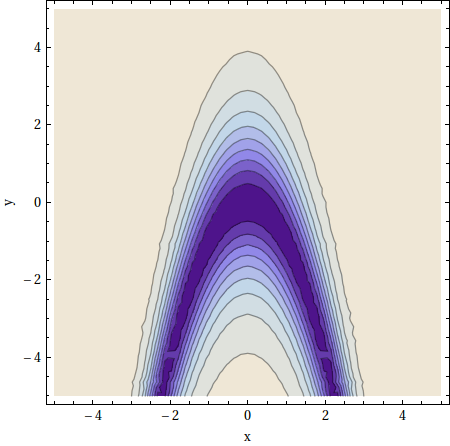}
\ \
\includegraphics[width=7
cm]{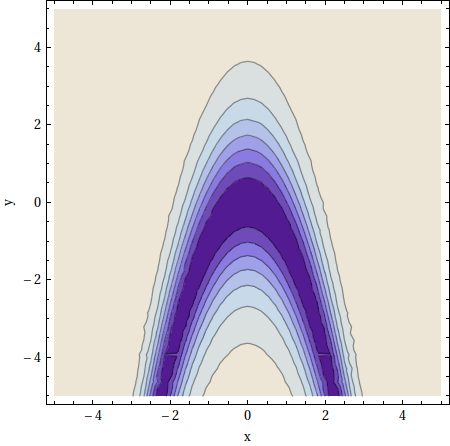}
\caption{The contour plots of the curved dark solitons $u$ and $v$ given by (\ref{csolfinal}) in $x-y$ plane at $t=0$ for the choice of $B(X) = X^2 = (x - a t)^2.$ The constants are chosen as $m_1 = 1, m_2 = 2, a = P = 1, |\tau_1|^2 = 1/10, |\tau_2|^2 = 1/2.$ Due to the presence of nonlinear function $B(X)$ in the argument, the solitons  (\ref{csolfinal}) get bend in $x-y$ plane.  } 

\label{NDTEplot}
\end{figure}

\begin{figure}[hbt!]
\centering
\includegraphics[width=7cm]{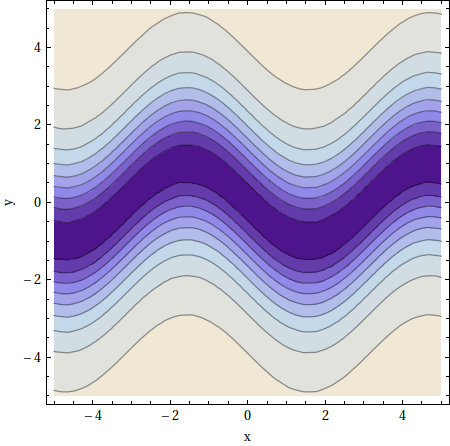}
\ \
\includegraphics[width=7
cm]{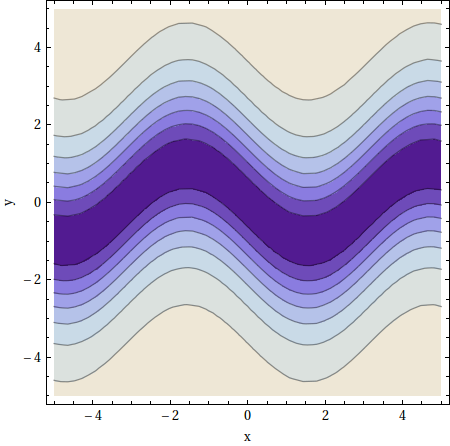}
\caption{The contour plots of the curved dark solitons $u$ and $v$ given by (\ref{csolfinal}) in $x-y$ plane at $t=0$ for the choice of $B(X) = \sin{X} = \sin{(x - a t)}.$ The constants are chosen as $m_1 = 1, m_2 = 2, a = P = 1, |\tau_1|^2 = 1/10, |\tau_2|^2 = 1/2.$ Due to the presence of nonlinear function $B(X)$ in the argument, the solitons  (\ref{csolfinal}) get bend in $x-y$ plane.} 

\label{NDTEplot}
\end{figure}

\begin{figure}[hbt!]
\centering
\includegraphics[width=7cm]{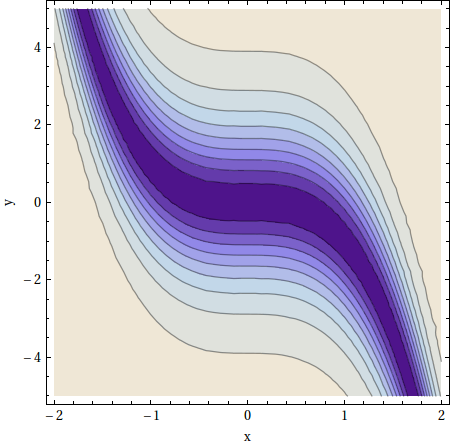}
\ \
\includegraphics[width=7
cm]{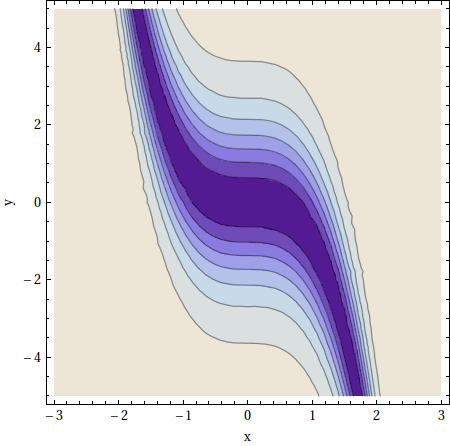}
\caption{The contour plots of the curved dark solitons $u$ and $v$ given by (\ref{csolfinal}) in $x-y$ plane at $t=0$ for the choice of $B(X) = X^3 = (x - a t)^3.$ The constants are chosen as $m_1 = 1, m_2 = 2, a = P = 1, |\tau_1|^2 = 1/10, |\tau_2|^2 = 1/2.$ Due to the presence of nonlinear function $B(X)$ in the argument, the solitons  (\ref{csolfinal}) get bend in $x-y$ plane.} 

\label{NDTEplot}
\end{figure}

Thus we can form different shapes of the dark soliton solutions (\ref{csolfinal}) by choosing different functional forms of the arbitrary function $B(X), X = x - a t$ present in the argument of the $tanh$  function.
This property may  be useful in modeling the bending of light beam \cite{KunduNaskar} in nonlinear optics. Previously, the bending of an optical  beam
was shown  in \cite{Bending1,Bending2} through the Airy function solution of linear Schrodinger equation.  It can preserve its parabolic form over a
finite distance. The beam acceleration along an arbitrary curve was shown in \cite{KunduNaskar} by Kundu and Naskar where the shape was not preserved. Such bending is controlled by the boundary population inversion of atoms in an Erbium doped medium.
In that situation, the boundary value of population inversion of the medium becomes non uniform.
On the other hand, the bending feature of this work is more general and arbitrary. Due to the presence of the arbitrary function $B(X)$, any shape can be obtained which is important for Mathematical modelling of experimental data. Also, the amplitude of the soliton remains constant. Such property, that is the existence of the arbitrary function $B(X)$ causing arbitrary bending, originates from the current like nonlinearity and Galilean Co-variance of the coupled KMN system (\ref{cKMN}). It is a very rare property which is absent in NLSE with Kerr like nonlinearity.

Now it is important to
discuss the condition for larger bending, it would be interesting to derive the amount of  curvature due to function $B(X)$.
 For the static case
($t = 0$), the locus of the maximum amplitude of the curved
soliton solution (\ref{csolfinal}) is of the form 
\begin{equation}
 B(x) + P y + \xi_1^0 = 0.
\end{equation}
Now taking the derivative with respect to $x$ twice, we get 
\begin{equation}
 \frac{dS}{dx} = - \frac{d^2B}{dx^2}.\label{curvature} \end{equation}
where the slope is defined as $S = \frac{dy}{dx}$. We see from (\ref{curvature}) that
for a higher value of $- \frac{d^2B}{dx^2}$, the rate of variation of
the slope $S$ is higher. Hence, the slope of the
maximum amplitude curve will vary large for traveling the
unit distance in $x$ . Larger rate of variation of  $S$ describes
larger bending. Hence, for large bending to
take place, $- \frac{d^2B}{dx^2}$ should be high which can be seen from FIG. 3,4,5 respectively.

\section{Conclusive remarks}
In this work, we have considered the coupled Kundu Mukherjee Naskar system which was introduced by Yildirim in 
\cite{Optik-2KMN-2019, Optik-2KMN-2019a} to model (2+1) dimensional optical solitons in birefringence fiber without four wave mixing terms. The Galilean co-variance property of the system has been derived. The dark and bright one solitons of the system has been calculated using Hirota bilinearization method. It has been shown that the dark solitons of the two modes $u$ and $v$ can
   bend arbitrarily due to the presence of an arbitrary function of space (x) and time (t) which is shown by the contour plots. Such arbitrary curvature originates from the current like nonlinearity and Galilean co-variance present in the system (\ref{cKMN}). The condition of larger bending has been derived and explained. Such analytical solutions  may pave new directions of research in the  field of Nonlinear Optics.

\section {ACKNOWLEDGMENTS}

The work is sponsored by Shanghai Pujiang Program and Tsung Dao Lee Institute starting grant.
% The work has been carried out with partial financial support from the
% Ministry of Science and Higher Education of the Russian Federation in the
% framework of Increase Competitiveness Program of NUST MISiS $( K4-2018-061),$ implemented by a governmental decree dated 16 th of March 2013, N 211.
Author is indebted to Ms. M. Mukherjee for the thoughtful discussions and unconditional support during the progress of the work.

\end{document}